\documentclass[%
 reprint,
nofootinbib,
 amsmath,amssymb,
 aps,prd,
]{revtex4-1}

\usepackage[sort&compress]{natbib}
\usepackage{color}
\usepackage{dblfnote}
\DFNalwaysdouble
\usepackage{slashed}
\usepackage{graphicx}
\usepackage{dcolumn}
\usepackage{bm}
\usepackage{hyperref}

\newcommand{\mrm}[1]{_{\rm #1}}
\renewcommand{\d}{{\rm d}}
\newcommand{\citeeq}[1]{Eq.~(\ref{#1})}

\newcommand{\citepeq}[1]{Eq.~(\ref{#1})}
\newcommand{\citepfig}[1]{Fig.~\ref{#1}}

\begin{document}
%
%

\title{Evolution of primordial black hole spin due to Hawking radiation}

\author{Alexandre Arbey}
 \altaffiliation[Also at ]{Institut Universitaire de France, 103 boulevard Saint-Michel, 75005 Paris, France}%
 \email{alexandre.arbey@ens-lyon.fr}
\affiliation{Univ Lyon, Univ Claude Bernard Lyon 1, CNRS/IN2P3, IP2I Lyon, UMR 5822, F-69622, Villeurbanne, France}

\author{J\'er\'emy Auffinger}
\email{j.auffinger@ipnl.in2p3.fr}
\affiliation{
Institut d'Astrophysique de Paris, UMR 7095 CNRS, Sorbonne Universit\'es,
98 bis, boulevard Arago, F-75014, Paris, France \\
 Univ Lyon, Univ Claude Bernard Lyon 1, CNRS/IN2P3, IP2I Lyon, UMR 5822, F-69622, Villeurbanne, France \\
  D\'epartement de Physique, \'Ecole Normale Sup\'erieure de Lyon, F-69342 Lyon, France
}

\author{Joseph Silk}
\email{joseph.silk@physics.ox.ac.uk}
\affiliation{%
 Institut d'Astrophysique de Paris, UMR 7095 CNRS, 
 Sorbonne Universit\'es,
98 bis, boulevard Arago, F-75014, Paris, France \\
The Johns Hopkins University, Department of Physics and Astronomy, Baltimore, Maryland 21218, USA\\
Beecroft Institute of Particle Astrophysics and Cosmology, University of Oxford, Oxford OX1 3RH, UK
}

\date{\today}

\begin{abstract}
	Near extremal Kerr black holes are subject to the Thorne limit $a<a^*\mrm{lim}=0.998$ in the case of thin disc accretion, or some generalized version of this in other disc geometries. However any limit that differs from the thermodynamics limit $a^* < 1$ can in principle be evaded in other astrophysical configurations, and in particular if the near extremal black holes are primordial and subject to evaporation by Hawking radiation only. We derive the lower mass limit above which Hawking radiation is slow enough so that a primordial black hole with a spin initially above some generalized Thorne limit can still be above this limit today. Thus, we point out that the observation of Kerr black holes with extremely high spin should be a hint of either exotic astrophysical mechanisms or primordial origin.
\end{abstract}

\pacs{Valid PACS appear here}
\maketitle


\section{\label{sec:intro}Introduction}

Primordial Black Holes (PBHs) are appealing candidates for solving the long-standing issue of dark matter \citep{Carr2010,Carr2016,GarciaBellido2017,Clesse2018}. PBHs could have been created at the end of the inflationary stage of the early Universe, when relatively high density fluctuations $\Delta\rho/\rho \gtrsim 1$ re-entered the Hubble horizon. The mass collapsing into a PBH through this mechanism is not subject to the lower Chandrasekhar limit of $\sim 1.4\,M_\odot$ \citep{Hawking1989book}, as this limit is a consequence of the stellar origin of  Black Holes (BHs). Thus, the detection of a sub-solar BH in the merger events of forthcoming gravitational waves detectors such as LISA \citep{LISA2013} would certainly point to a primordial origin \citep{Chen2019}.

Most excitingly, there are powerful observational constraints, primarily from gravitational microlensing in the subsolar mass regime, but a substantial window remains open for PBHs as dark matter in the mass range that extends from asteroid mass scales down to the mass set by evaporation limits \citep{Niikura2019}.

In principle, depending on their mechanism of production at the end of inflation, there is no restriction on the initial spin of a PBH up to near extremal values $a^* \lesssim 1$. On the other hand, for BHs with astrophysical origin, Thorne has shown that in the case of thin disc accretion, there is a limit to the reduced spin $a^*\mrm{lim} \approx 0.998$. This limit comes from accretion of the surrounding gas on a BH, and its balance with superradiance effects \citep{Thorne1974}. Surprisingly, the same $a^*\mrm{lim} \approx 0.998$ is found in \citep{Kesden2010} for BH mergers. In the case of accretion, this limit has been recently generalized to other accretion regimes and disc geometries in \citep{Skadowsy2011}, based on earlier work \citep{Abramowicz1980}; reaching somewhat higher values depending on the disc parameters. The overall state-of-art is that, except for really specific accretion environments, the spin of a BH of astrophysical origin should not exceed a generalized version of the Thorne limit \citep{Nampalliwar2018}. However, the superior limit $a^* < 1$ holds in any case due to the third law of thermodynamics \citep{Bardeen1973}. Indeed, a BH with $a^* = 1$ would have $T = 0$, which is classically forbidden for any statistical system. Moreover, its horizon would have disappeared, thus revealing a naked space-time singularity and violating the Cosmic Censorship Conjecture (a comprehensive discussion of extremal BHs is given in e.g. \cite{Belgiorno2004,Lehman2019}). We would like to emphasize the fact that, even if the thermodynamics limit $a^* < 1$ remains true, {\it nothing prevents a priori astrophysical BHs from reaching a near extremal spin value $a^* = 0.999999...$, yet the mechanisms are still to be proposed.}

Thus, the detection of BHs with a reduced spin higher than a generalized Thorne limit $a^* > a^*\mrm{lim}$ could point either towards primordial origin \citep{Taylor1998,Fernandez2019} or astrophysical origin with an exotic accretion history.

In this paper, we focus on the mechanisms allowing a PBH to have today a spin higher than a generalized Thorne limit. For this purpose, we compute the mass and spin loss through Hawking radiation of a Kerr PBH and evaluate the minimum initial mass a PBH should have in order to experience a current spin value above a generalized Thorne limit.

\section{\label{sec:Hawking}Hawking radiation}

\subsection{\label{sec:theory}Theoretical aspects}

\subsubsection{Emission rate}

Hawking showed that BHs are not as black as was first supposed \citep{Hawking1975}. Throughout,  we use a natural system of units where $G = c = k\mrm{B} = \hbar = 1$. Hawking used a semi-classical treatment, that is to say the general relativity Kerr (or Schwarzschild) metric for space-time
\begin{align}
\d s^2 &= \left( 1 - \dfrac{2Mr}{\Sigma^2} \right)\d t^2 + \dfrac{4aMr\sin(\theta)^2}{\Sigma^2}\d t\d \phi - \dfrac{\Sigma^2}{\Delta} \d r^2 \nonumber\\
&- \Sigma^2\d\theta^2 - \left( r^2 + a^2 + \dfrac{2a^2Mr\sin(\theta)^2}{\Sigma^2} \right)\sin(\theta)^2\d\phi^2\,, \label{eq:kerr}
\end{align}
where $M$ is the BH mass, $a \equiv J/M$ is the BH spin parameter ($J$ is the BH angular momentum), $\Sigma \equiv  r^2 + a^2\cos(\theta)^2$ and $\Delta \equiv r^2 -2Mr + a^2$, and a quantum mechanics treatment of Standard Model (SM) particles through a wave function $\psi$ satisfying the Dirac equation for fermions (spin $1/2$)
\begin{equation}
(i\slashed{\partial} - \mu)\psi = 0\,, \label{eq:dirac}
\end{equation}
where $\slashed{\partial} \equiv \gamma_\nu \partial^\nu$ is the standard Feynman notation, and the Proca equation for bosons (spin 0, 1 or 2)
\begin{equation}
(\square + \mu^2)\psi = 0\,, \label{eq:proca}
\end{equation}
where $\mu$ is the particle rest mass. Setting $\mu = 0$ in these equations of motion also allows to compute the propagation of the massless fields (in the following neutrinos, photons, gravitons). In these equations, we neglect the couplings between the fields, since they do not affect the probability of emission of (primary) SM particles via Hawking radiation, but we consider them to obtain the abundance of the final (secondary) particles at infinity, which come from the hadronization or decay of the primary particles. Solving these equations shows that there is a net emission of particles of type $i$ called the Hawking radiation (HR). The number of particles emitted per unit time and energy is
\begin{equation}
\dfrac{\d^2N_i}{\d t\d E} = \dfrac{1}{2\pi}\sum_{\rm dof.}\dfrac{\Gamma^{lm}_i(E,M,a^*)}{e^{E^\prime/T}\pm 1}\,, \label{eq:hawking}
\end{equation}
where $T$ is the Kerr BH Hawking temperature
\begin{equation}
T \equiv \dfrac{1}{2\pi}\left( \dfrac{r_+ - M}{r_+^2 + a^2} \right)\,,
\end{equation}
and $r_\pm \equiv M\left(1 \pm \sqrt{1 - (a^{*})^2}\right)$ are the Kerr horizons radii; $a^* \equiv a/M$ is the Kerr dimensionless spin parameter, it is 0 for a Schwarzschild -- non rotating -- BH and 1 for a Kerr extremal BH; $E^\prime \equiv E - m\Omega$ is the energy of the particle that takes into account the horizon rotation with angular velocity $\Omega \equiv a^*/(2r_+)$ on top of the total energy $E \equiv E\mrm{kin.} + \mu$; $m$ is the particle angular momentum projection $m \in [-l,+l]$. The sum of \citepeq{eq:hawking} is on the degrees of freedom (dof.) of the particle considered, that is to say the color and helicity multiplicity as well as the angular momentum $l$ and its projection $m$. The quantity $\Gamma_i^{lm}(E,M,a^*)$ is called the greybody factor and has been extensively studied in the literature (see below). It encodes the probability that a particle of type $i$ with angular momentum $l$ and projection $m$ generated at the horizon of a BH escapes its gravitational well and reaches space infinity.

\subsubsection{Evolution of BHs}

After computing the greybody factors $\Gamma_i^{lm}$, it is possible to compute the mass and spin loss rates by integrating \citepeq{eq:hawking} over all energies and summing over all -- massive and massless -- SM particles $i$ (6 quarks + 6 antiquarks, 3 neutrinos + 3 antineutrinos, 3 charged leptons + 3 charged antileptons, 8 gluons, weak $W^+$, $W^-$ and $Z^0$ bosons, the photon and the Higgs boson), plus the graviton. We define the (positive) $f$ and $g$ factors following \citep{Page1976,Dong2016}
\begin{align}
f(M,a^*) &\equiv -M^2 \dfrac{\d M}{\d t}\label{eq:fM} \\
&= M^2\int_{0}^{+\infty} \sum_i\sum_{\rm dof.} \dfrac{E}{2\pi}\dfrac{\Gamma_i^{lm}(E,M,a^*)}{e^{E^\prime/T}\pm 1} \d E \,, \nonumber 
\end{align}
\begin{align}
g(M,a^*) &\equiv -\dfrac{M}{a^*} \dfrac{\d J}{\d t}\label{eq:gM} \\ 
&= \dfrac{M}{a^*}\int_{0}^{+\infty} \sum_i\sum_{\rm dof.}\dfrac{m}{2\pi} \dfrac{\Gamma_i^{lm}(E,M,a^*)}{e^{E^\prime/T}\pm 1}\d E \,. \nonumber 
\end{align}
Inverting these equations and using the definition of $a^*$, we obtain the differential equations governing the mass and spin of a Kerr BH
\begin{equation}
\dfrac{\d M}{\d t} = -\dfrac{f(M,a^*)}{M^2}\,, \label{eq:diffM}
\end{equation}
and
\begin{equation}
\dfrac{\d a^*}{\d t} = \dfrac{a^*(2f(M,a^*) - g(M,a^*))}{M^3}\,. \label{eq:diffa}
\end{equation}

\subsection{\label{sec:numerical}Numerical implementation}

We solve Eqs.~\eqref{eq:diffM} and \eqref{eq:diffa} numerically, using a new code entitled \texttt{BlackHawk} \citep{BlackHawk}\footnote{\url{https://blackhawk.hepforge.org/}}. This code contains tabulated values of $f(M,a^*)$ and $g(M,a^*)$ obtained through Eqs.~\eqref{eq:fM} and \eqref{eq:gM}. Within \texttt{BlackHawk}, efforts have been made to compute the greybody factors $\Gamma^{lm}_i(E,M,a^*)$ numerically.

Teukolsky and Press \citep{Teukolsky1,Teukolsky2} have shown that the Dirac and Proca equations \eqref{eq:dirac} and \eqref{eq:proca}, once developed in the Kerr metric \eqref{eq:kerr}, can be separated into a radial and an angular part with the wave function written as $\psi(t,r,\theta,\phi) = R(r) \, S_{lm}(\theta) \, e^{-iEt} \,e^{im\phi}$. In the following, we make the approximation that the particles are massless for the greybody factor computations, since the main effect of the non-zero particle rest mass is to induce a cut in the emission spectra at energies $E < \mu$ \citep{Page1977}. This cut in the energy spectra is applied as a post-process and included in the computation of the integrals \eqref{eq:fM} and \eqref{eq:gM}. We checked that the approximated spectra only slightly differ from the ones with a full massive computation, and the small differences are smoothed out by the integration in Eqs.~(\ref{eq:fM}) and (\ref{eq:gM}). The radial equation on $R(r)$ for a massless field of spin $s$ reads
\begin{align}
	\frac{1}{\Delta^s}\frac{\d}{\d r}&\left( \Delta^{s+1}\frac{\d R}{\d r} \right)\nonumber\\ + &\bigg( \dfrac{K^2+2is(r-M)K}{\Delta} - 4is E r - \lambda_{slm}\bigg)R = 0\,, \label{eq:teukolsky}
\end{align}
where $\lambda_{slm}$ is the eigenvalue of the angular part (we use the polynomial expansion of \citet{Dong2016} to compute $\lambda_{slm}$) and $K\equiv (r^2 + a^{*2}M^2)E + a^*Mm$. Then, Chandrasekhar and Detweiler \citep{Chandra1,Chandra2,Chandra3,Chandra4} have shown that through suitable function transformation $R\rightarrow Z$ and a change of variable from the Boyet-Lindquist radial coordinate to a generalized tortoise coordinate $r\rightarrow r^*$, one can transform \citepeq{eq:teukolsky} into a wave equation with a short-range potential
\begin{equation}
\dfrac{\d^2 Z}{\d r^{*2}} + \left(E^2 - V(r^*)\right)Z = 0\,. \label{eq:schrodinger}
\end{equation}
For details about this transformation, we refer the reader to Appendix~\ref{app:change_of_var}. We solve this wave equation numerically with \texttt{Mathematica}, starting from an ingoing plane wave at the horizon
\begin{equation}
Z\mrm{hor} \underset{r^*\rightarrow -\infty}{\rightarrow} e^{iEr^*}\,,
\end{equation}
and integrating to space infinity where the solution is
\begin{equation}
Z_\infty \underset{r^*\rightarrow +\infty}{\rightarrow} Ae^{iEr^*} + Be^{-iEr^*}\,,
\end{equation}
we identify the transmission coefficient (greybody factor)
\begin{equation}
\Gamma \equiv |A|^2\,.
\end{equation}
This allows us to perform the integrals \eqref{eq:fM} and \eqref{eq:gM}.

\texttt{BlackHawk} uses an Euler-based adaptive time step method to compute accurately the last stages of the BH life, when its mass goes down to the Planck mass $M\mrm{P}$ very quickly. In practice, the time step is decreased whenever the relative changes in mass or spin are too large ($|\Delta M|/M$ or $|\Delta a^*|/a^*$ $>0.1$), leading to a precision of a few percent. When $M\sim M\mrm{P}$, we consider that Hawking evaporation terminates.

\section{Results}

\subsection{Evolution of Kerr BHs}

The main difference between Kerr ($a^* \ne 0$) and Schwarzschild ($a^* = 0$) BHs is that Kerr BHs are axially symmetric and not spherically symmetric. This gives a favored axial direction when computing the Hawking radiation. The emission of particles with an angular momentum spinning in the same direction as that of the BH is enhanced when $a^*$ increases. Moreover, for sufficiently small energies and high angular momentum
\begin{equation}
E < E\mrm{SR} \equiv \dfrac{a^*m}{2r_+}\,,
\end{equation}
we enter the regime of superradiance (SR), with enhanced emission. This asymmetry in the Hawking radiation causes a net spin loss by the BH (hence the positivity of the $g$ factor defined in \citepeq{eq:gM}) through the emission of high angular momentum particles. This enhanced radiation also causes a mass loss larger than in the Schwarzschild case. Thus, Kerr BHs have a shorter lifetime than Schwarzschild BHs, and it gets shorter and shorter as the initial spin $a^*_i$ gets close to 1.

\begin{figure}
	\centering
	\includegraphics[width = 0.45\textwidth]{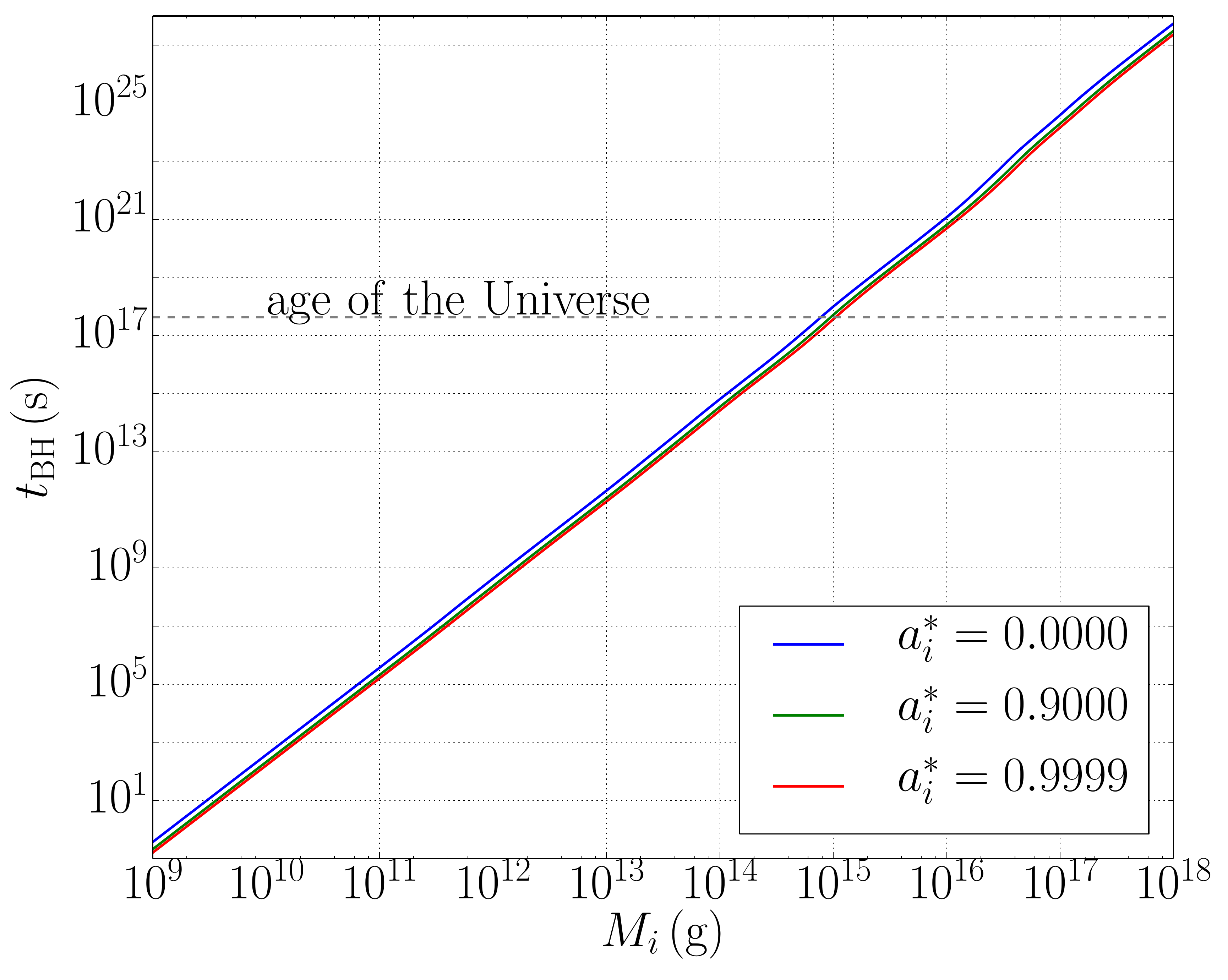}
	\caption{Kerr BH lifetimes $t\mrm{BH}$ as functions of the BH mass $M\mrm{BH}$ for different initial spins $a^*_i = \{0, 0.9, 0.9999\}$ (blue, green and red curves respectively). The age of the universe is indicated as a grey horizontal line.}
	\label{fig:lifetimes}
\end{figure}

Fig.~\ref{fig:lifetimes} shows the lifetime of Kerr BHs as a function of their mass for different initial spins $a^*_i$. The Hawking radiation computed with \texttt{BlackHawk} for the evaporation includes all the SM particles (both massive and massless) as well as one massless graviton. We see that the spin indeed reduces the lifetime but the difference is small compared to the enormous time range spanned by the BH lifetimes. We verify that the lifetime is approximately given by
\begin{equation}
t\mrm{BH} \sim M^3\,,
\end{equation}
which can be derived from \citepeq{eq:diffM} by considering that $f(M,a^*)$ is a constant. This approximate relation still holds in presence of angular momentum $a^* \ne 0$.

\begin{figure}
	\centering{
		\includegraphics[width = 0.45\textwidth]{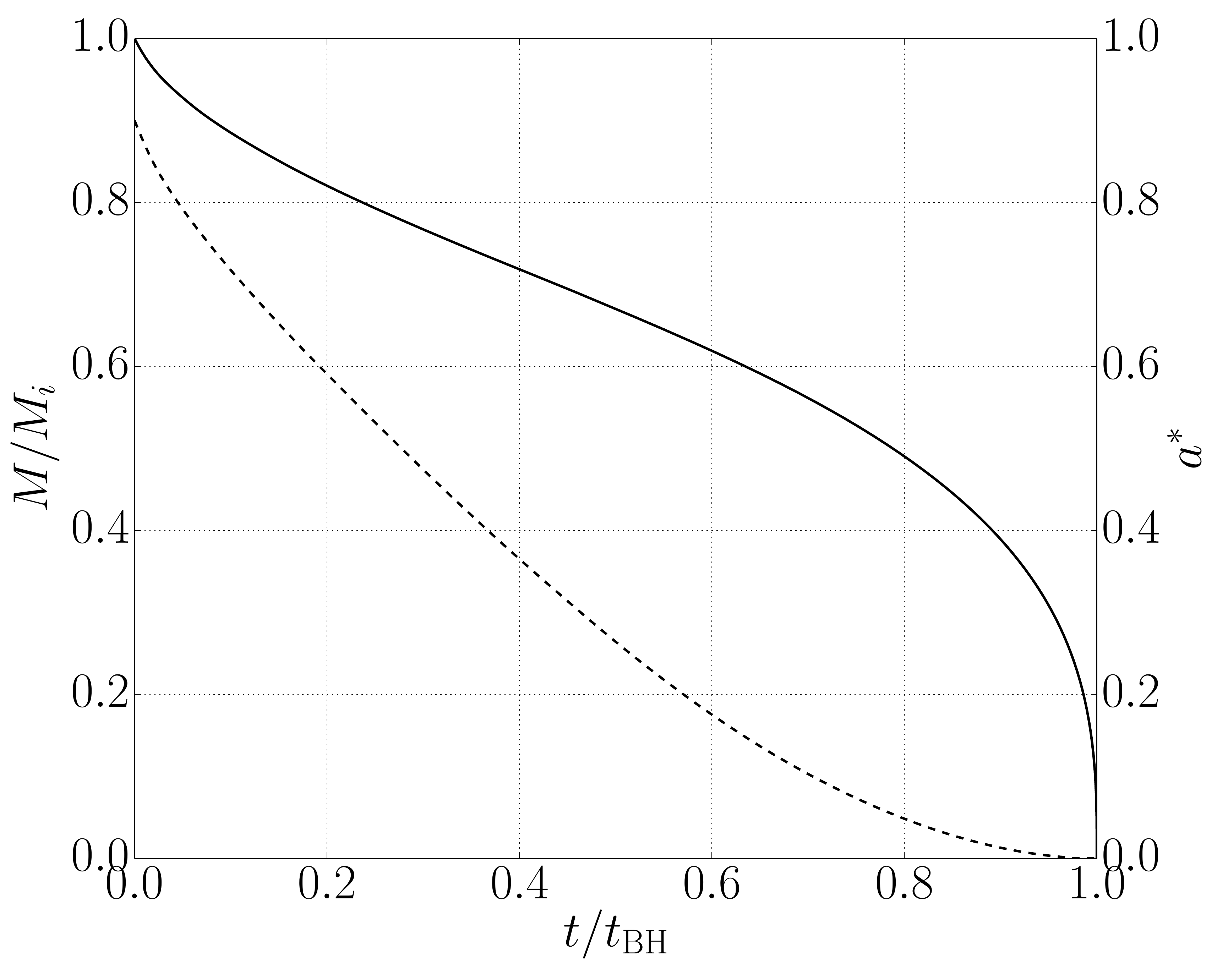}
		\caption{Kerr BH mass $M$ (plain curve, normalized over the initial mass $M_i = 10^{16}\,$g) and reduced spin $a^*$ (dashed curve, starting from an initial spin $a^*_i = 0.9$) as functions of time $t$ (normalized to the BH lifetime $t_{\rm BH}$).\label{fig:evolution_ex}}
	}
\end{figure}

\citepfig{fig:evolution_ex} shows an example of the evolution of the Kerr BH mass and spin through time.  We see that the reduced spin $a^*$ has a slightly shorter timescale than the mass $M$. This is easy to understand when looking at Eqs.~\eqref{eq:diffM} and \eqref{eq:diffa}. The first stage of the evolution is a strong decrease of both mass and spin, corresponding to the Kerr regime when the Hawking radiation is enhanced. When we leave the high-spin region ($a^*\lesssim 0.2$), the emission becomes similar to that of a Schwarzschild BH and the mass evolution is more monotonic. At the end of the BH life (the last $10\%$), a final stage of very fast evaporation occurs, during which the BH loses the major part of its mass ($\sim 50\%$). This is in agreement with the results of \citet{Taylor1998}. When reaching the Planck mass, Hawking's theory does not tell what happens of the BH.

\begin{figure}
	\centering{
		\includegraphics[width = 0.45\textwidth]{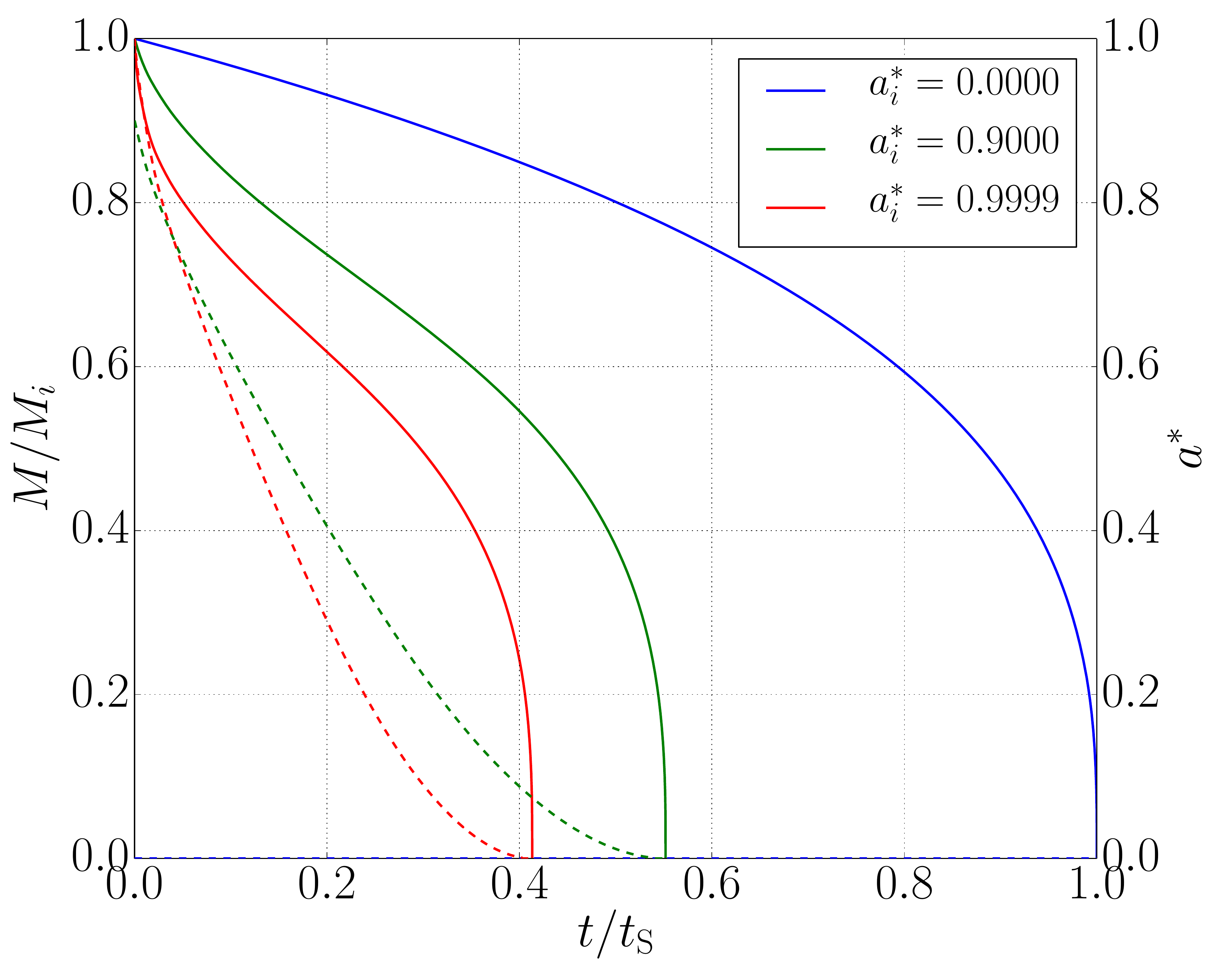}
		\caption{Comparison of Kerr BH mass $M$ (plain curves, normalized to the initial mass $M_i = 10^{16}\,$g which is the same for all curves) and spin $a^*$ (dashed curves) evolutions as functions of time $t$ (normalized to the Schwarzschild BH lifetime $t_{\rm S}$), for different values of the initial spin $a^*_i$ ranging (right to left) from $a_i^* = 0$ (Schwarzschild case) to $a_i^* = 0.9999$ (near extremal case).\label{fig:evolutions}}
	}
\end{figure}

\citepfig{fig:evolutions} shows the mass and spin evolutions for the same initial mass $M_i = 10^{16}\,$g but different initial spins $a_i^* = \{0, 0.9, 0.9999\}$. We see that the lifetime of a Kerr BH can be reduced by almost $\sim60\%$ when going from the Schwarzschild case $a^*_i = 0$ to the near extremal case $a^*_i = 0.9999$. This is compatible with the results of \citet{Dong2016}. The higher the initial spin is, the stronger the initial mass loss will be, so the shorter the BH lifetime. We can see that after most of the spin is radiated away, all curves share the same shape as the Schwarzschild one.

\begin{figure}
	\centering{
		\includegraphics[width = 0.45\textwidth]{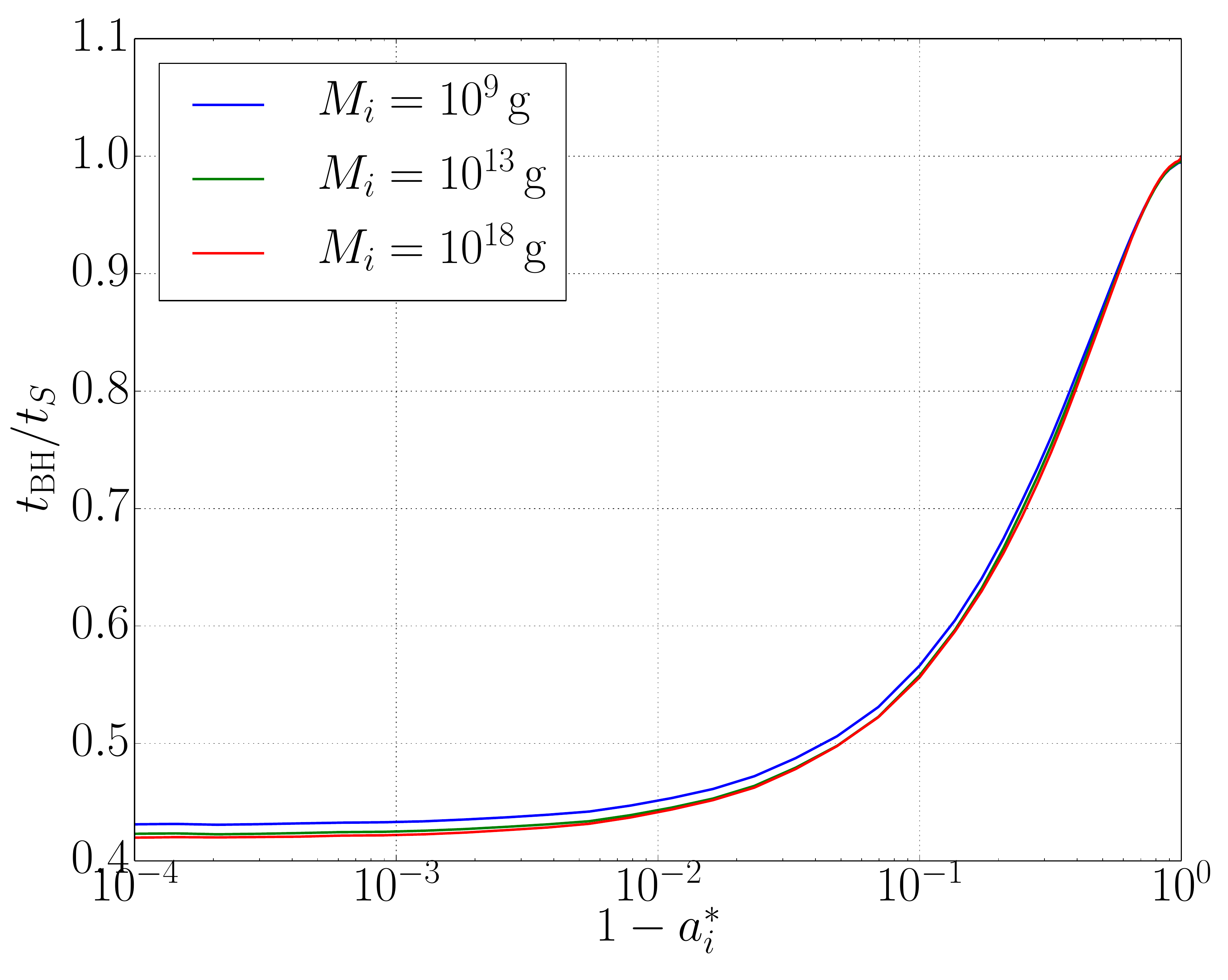}
		\caption{Kerr BH lifetimes $t\mrm{BH}$ (normalized to the Schwarzschild case $t_{\rm S}$) for different initial masses $M_i = \{10^{9}, 10^{13}, 10^{18}\}\,$g (blue, green and red lines, respectively) as functions of the initial spin $a^*_i$. The $x$-axis has been reversed to show $1-a^*_i$ in a logarithmic scale.\label{fig:life_times}}
	}
\end{figure}

\citepfig{fig:life_times} shows the evolution of the lifetime of a Kerr BH as a function of the initial spin $a^*_i$, for different values of the initial mass $M_i = \{10^{9}, 10^{13}, 10^{18}\}\,$g. We have reversed the $x$-axis to focus on the near extremal region $a^*_i \lesssim 1$. We see that the lifetime decreases as the initial spin increases, but this saturates as we come closer to the extremal Kerr case $a^*_i \lesssim 1$. The decrease of the lifetime relative to the Schwarzschild case is not the same for all initial masses since they have a different Hawking emission history: lighter BHs can emit massive particles at the beginning of their evaporation (in the Kerr regime) while heavier BHs can only emit them at the end of their evaporation (in the Schwarzschild regime). The difference in the evolution of the lifetimes remains small.

\subsection{Maximum spin}

Using these data on the Kerr BH evolution, which is a function of both mass and spin, we can estimate the maximum spin a BH can still have today, starting from some initial spin, and depending on its initial mass. We know that some generalized Thorne limit prevents BHs with a disc from having a spin higher than $a^*\mrm{lim} \gtrsim 0.998$, due to accretion and superradiance effects \citep{Thorne1974,Abramowicz1980,Skadowsy2011}. We also know that the same limit applies to the outcome of BH mergers due to general relativistic dynamics \citep{Kesden2010}. One possibility of overcoming $a^*\mrm{lim}$, while remaining below the thermodynamics limit $a^* < 1$, may be to form a Kerr PBH with an initial spin $a_i^* > a^*\mrm{lim}$ and to maintain this spin over time until today. As mentioned in the Introduction, the precise value of the Thorne limit can depend on the disc geometry and parameters (accretion regime, viscosity) \citep{Skadowsy2011}, thus the numerical results presented in this Section have to be adapted to somewhat higher generalized Thorne limits.

\begin{figure}
	\centering{
		\includegraphics[width = 0.45\textwidth]{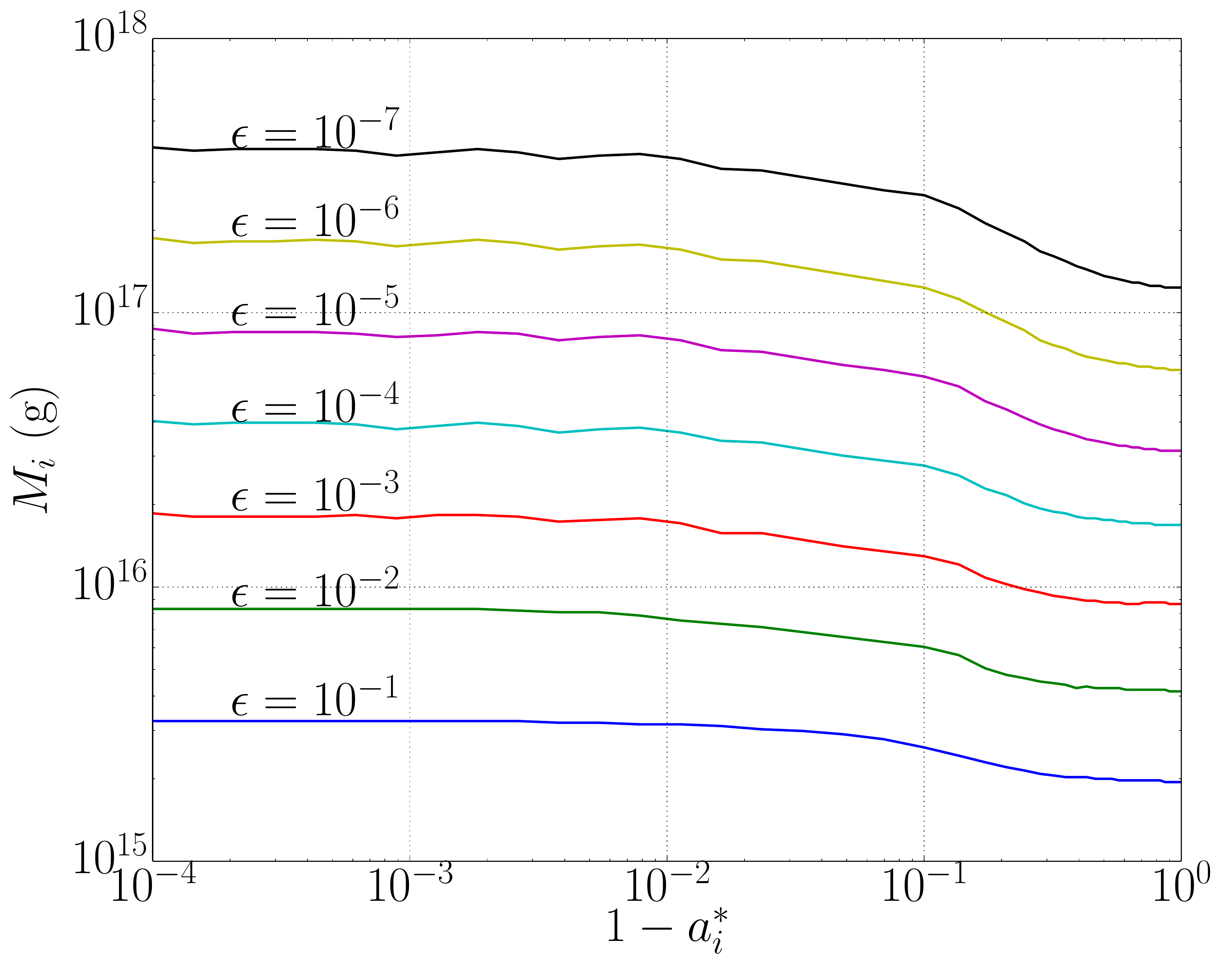}
		\caption{Minimum initial mass $M_i$ needed to have a relative spin loss today $\epsilon \equiv \Delta a^*/a^*_i$ for different values of $\epsilon$. The $x$-axis has been reversed to show $1-a^*_i$ in a logarithmic scale.\label{fig:unchanged_spin}}
	}
\end{figure}

We have seen that the spin decrease time-scale corresponds roughly to that of the mass decrease $t\mrm{BH} \sim M_i^3$. That means that in order to maintain a spin value really close to the extremal Kerr case, the BH initial mass must be sufficiently high. \citepfig{fig:unchanged_spin} shows the minimum initial mass needed as a function of the initial spin, for different values of the desired relative spin change $\epsilon \equiv \Delta a^* / a^*_i$. As expected, the more we want to have a spin today close to the initial one ($\epsilon \rightarrow 0$) the more massive the BH has to have been originally. As $\epsilon\rightarrow 1$ (all initial spin is lost), the minimum mass, for all initial spins, goes to $M\mrm{lim}(a^*_i) \sim 10^{15}\,$g the mass of the BHs just evaporating  today. 

\begin{figure}
	\centering{
		\includegraphics[width = 0.45\textwidth]{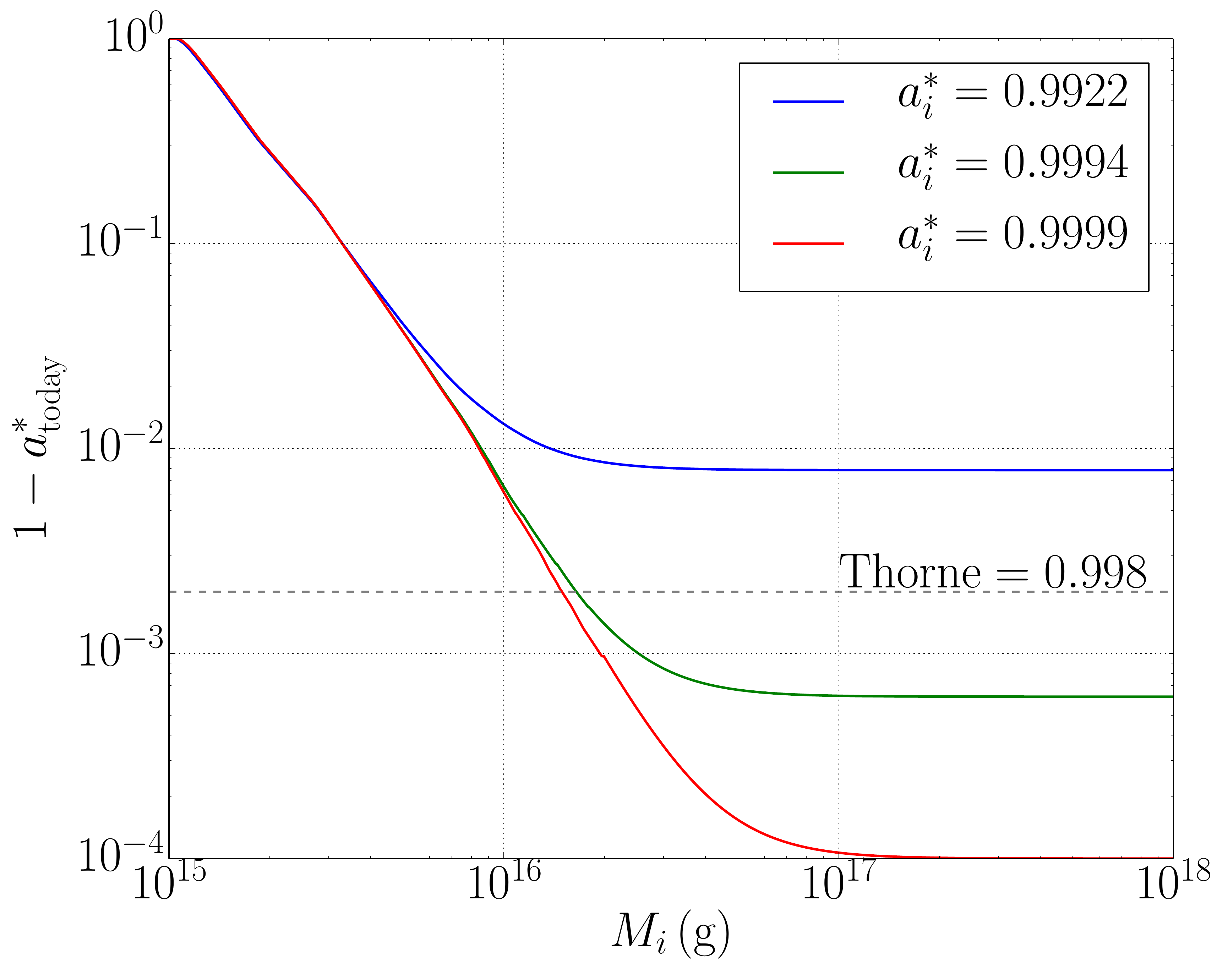}
		\caption{Value of the spin today $a^*_{\rm today}$ as a function of the initial mass $M_i$ for different initial spins $a^*_i = \{0.9922, 0.9994, 0.9999\}$ (blue, green and red lines, respectively). The Thorne limit $a^*\mrm{lim} \approx 0.998$ is shown as an horizontal dashed line. The $y$-axis has been reversed to show $1-a^*_{\rm today}$ on a logarithmic scale.\label{fig:today_spin}}
	}
\end{figure}

\citepfig{fig:today_spin} gives a reversed view of \citepfig{fig:unchanged_spin}: starting from an initial spin $a^*_i = \{0.9922, 0.9994, 0.9999\}$ below or above the Thorne limit, it shows the value of the spin today $a^*_0$ as a function of the initial mass. We see that for sufficiently high initial masses $M_i \gtrsim 10^{16}-10^{17}\,$g, the value of the spin has barely changed, as could be already guessed from \citepfig{fig:unchanged_spin}. For initial masses $M_i > M_i^{\rm lim} \approx 2-3\times 10^{16}\,$g, the spin value today is still higher than the Thorne limit for the two cases where it was higher at the beginning. That means that a Kerr PBH of initial spin $a^*_i > a^*\mrm{lim}$ could still have a spin $a^*_0 > a^*\mrm{lim}$ today if it were sufficiently heavy. The same picture could be drawn for even higher initial spins $a^*_i = 0.999999...$ with a decrease of $M_i^{\rm lim}$ when $a^*_i$ increases. Indeed, starting from a higher spin, a smaller initial mass is necessary to reach the Thorne limit today through Hawking radiation.

\subsection{Accretion and mergers}

The above computation of the PBH spin evolution is relevant only if the mechanisms leading to the establishment of the generalized Thorne limit are avoided, that is to say accretion of material surrounding the PBH and mergers with other PBHs. The accretion part is clearly not a problem as accretion is dominated by Hawking radiation for sufficiently light PBHs during the radiation-dominated era. During the matter-dominated era, PBHs do not necessarily evolve in a matter-rich environment as they do not come from the collapse of a star. Thus, the spin loss is only given by the Hawking radiation, as computed with \texttt{BlackHawk}. The merging part should not be bothersome if the PBH merging rate is sufficiently small, which should be the case if PBHs do not contribute too much to the dark matter fraction (thus preventing the formation of binaries). At least, some of them should have been isolated until today. Thus, the generalized Thorne limit does not apply to sufficiently rare and light PBHs.

\subsection{Formation}

The question on how to generate such high-spin PBHs can be answered by a profusion of models of inflation and early Universe cosmology. Every model involving PBH creation should generate high spin PBHs at least as a tail in the distribution \citep{Chiba2017,Mirbabayi2019,DeLuca2019}. In these cases, the observation of high spin BHs should remain a rare event not necessarily incompatible with an unexpected astrophysical origin. However, some models predict a domination of high spin values for PBHs. We refer to one recent example, that of PBHs formed by scalar field fragmentation during the matter-dominated period that precedes reheating in an inflationary universe \citep{Cotner2017,Harada2017}. Precise spin measurements could be accomplished with further LIGO-Virgo data \citep{Fernandez2019}. We nevertheless point out that a more realistic scenario in which the transient matter-domination is not complete, taking into account the increasing proportion of radiation when this phase ends, predicts somewhat less extreme PBHs, and in fewer quantities \citep{Carr2018}. One last remark concerns the possible destruction of such extreme PBHs by external perturbations. It has been shown in \citet{Sorce2017} that falling matter can not overspin a near extremal Kerr-Newman BH, and a fortiori a near extremal Kerr one, thus the horizon should persist and the Hawking radiation paradigm should hold during the PBH evolution.

\section{Conclusions}

In this paper, we have computed the evolution of BH spin through Hawking radiation using our new code \texttt{BlackHawk}. We have seen that a way to presently have BHs with a spin near the Kerr extremal value $a^*\lesssim 1$ and above some generalized Thorne limit $a^*\mrm{lim} \gtrsim 0.998$ is to generate it in the primordial Universe through post-inflationary mechanisms with an initial spin $a^*_i > a^*\mrm{lim}$. Then, if its mass is sufficiently high, Hawking radiation is too slow to drive its spin below the generalized Thorne limit. One interesting result is that the initial PBH mass needed to retain such a high spin until today is well below the mass of the Sun.
We conclude that near extremal Kerr black holes may exist in nature, if primordial black holes constitute all of the dark matter in the observationally allowed window, or at least some of it in the higher mass range.
Moreover when such near extremal black holes enter the galactic environment, accretion of order 0.001 of their rest mass would render them sub-extremal and induce Hawking evaporation. Such potential black hole "bombs" may make primordial black holes directly detectable via X-ray or gamma ray emission.

\appendix

\section{From the Teukolsky equation to a Schr\"odinger-like wave equation}
\label{app:change_of_var}

In this Appendix we will briefly present the analytical method used by Chandrasekhar and Detweiler \cite{Chandra1,Chandra2,Chandra3,Chandra4} to transform the Teukolsky radial equation \citeeq{eq:teukolsky} into the Schr\"odinger-like wave equation \citeeq{eq:schrodinger}. The first change consists in moving from the Boyet-Lindquist coordinate $r$ to a generalized tortoise coordinate $r^*$ related to $r$ through
\begin{equation}
    \dfrac{{\rm d}r^*}{{\rm d}r} = \dfrac{\rho^2}{\Delta}\,, \label{eq:tortoise}
\end{equation}
where $\Delta \equiv r^2 - 2Mr + a^2$, $\rho \equiv r^2 + \alpha^2$ and $\alpha^2 \equiv a^2 + am/E$. This differential change of variable can be solved to give
\begin{align}
    r^*(r) = r &+ \dfrac{r\mrm{S}r_+ + am/E}{r_+ - r_-}\ln\left( \dfrac{r}{r_+} - 1 \right) \nonumber \\ 
    &- \dfrac{r\mrm{S}r_- + am/E}{r_+ - r_-}\ln\left( \dfrac{r}{r_-} - 1 \right)\,,
\end{align}
where $r\mrm{S} \equiv 2M$ is the Schwarzschild radius. The inverse relation has no analytical expression and must be computed numerically by solving the differential equation (\ref{eq:tortoise}). On top of this change, one changes the function from $R(r)$ to $Z(r^*)$ by imposing that the final result is a Schr\"odinger-like wave equation. Surprisingly, this is always something one can do for the values of the spin $s = 0,1,2,1/2$ and for both Schwarzschild ($a^* = 0$) and Kerr ($a^*\ne 0$) BHs. The precise transformations are given in the papers by Chandrasekhar and Detweiler and are of the form
\begin{align}
    &Z(r^*) = A(r^*)R(r(r^*)) + B(r^*)\dfrac{{\rm d}R}{{\rm d}r^*}\,, \\
    &\dfrac{{\rm d}Z}{{\rm d}r^*} = C(r^*)R(r(r^*)) + D(r^*)\dfrac{{\rm d}R}{{\rm d}r^*}\,.
\end{align}
Imposing that the equation governing $Z$ is \citeeq{eq:schrodinger} while $R$ satisfies \citeeq{eq:teukolsky} gives a system of equations that the functions $A$, $B$, $C$ and $D$ must fulfill. Solutions of this system give the form of the potential $V_s(r(r^*))$. These potentials are, for a field of spin $s$
\begin{equation}
	V_0(r) =\dfrac{\Delta}{\rho^4}\left( \lambda_{0\,lm} + \dfrac{\Delta + 2r(r-M)}{\rho^2} \dfrac{3r^2\Delta}{\rho^4} \right)\,, \label{eq:V0}
\end{equation}
\begin{align}
	V_{1/2,\pm}(r) =\, &(\lambda_{1/2\,lm}+1)\dfrac{\Delta}{\rho^4} \mp \dfrac{\sqrt{(\lambda_{1/2,l,m}+1)\Delta}}{\rho^4} \nonumber \\
	&\times\left( (r-M) - \dfrac{2r\Delta}{\rho^2} \right)\,, \label{eq:V12}
\end{align}
\begin{equation}
	V_{1,\pm}(r) = \dfrac{\Delta}{\rho^4}\left( (\lambda_{1\,lm}+2)-\alpha^2\dfrac{\Delta}{\rho^4} \mp i\alpha\rho^2 \dfrac{{\rm d}}{{\rm d}r}\left( \dfrac{\Delta}{\rho^4} \right) \right)\,, \label{eq:V1}
\end{equation}
\begin{align}
	V_{2}(r) = \,&\dfrac{\Delta}{\rho^8}\Bigg( q - \dfrac{\rho^2}{(q-\beta\Delta)^2}\Big( (q-\beta\Delta)\big( \rho^2\Delta q^{\prime\prime} \\ 
	& - 2\rho^2q - 2r(q^\prime\Delta - q\Delta^\prime)\big) + \rho^2\small(\kappa\rho^2 - q^\prime \nonumber\\
	& + \beta\Delta^\prime\small)(q^\prime\Delta - q\Delta^\prime) \Big) \Bigg)\,. \label{eq:V2}
\end{align}
The different potentials for a given spin lead to the same results. In the potential for spin 2 particles, the following quantities appear
\begin{align}
	&q(r) = \nu\rho^4 + 3\rho^2(r^2-a^2) - 3r^2\Delta\,, \\
	&q^\prime(r) = r\left( (4\nu + 6)\rho^2 - 6(r^2 - 3Mr + 2a^2) \right)\,, \\
	&q^{\prime\prime}(r) = (4\nu+6)\rho^2 + 8\nu r^2 - 6r^2 + 36Mr - 12a^2\,,
\end{align}
\begin{align}
	q^\prime\Delta - q\Delta^\prime = &-2(r-M)\nu\rho^4 + 2\rho^2(2\nu r\Delta -3M(r^2 + a^2)\nonumber \\
	&+ 6ra^2) + 12r\Delta(Mr - a^2)\,,
\end{align}
\begin{equation}
	\beta_\pm = \pm 3\alpha^2\,,
\end{equation}
\begin{equation}
	\kappa_\pm = \pm\sqrt{36M^2 -2\nu(\alpha^2(5\nu+6)-12a^2) + 2\beta\nu(\nu+2)}\,,
\end{equation}
\begin{equation}
	q-\beta_+\Delta = \rho^2(\nu\rho^2 + 6Mr - 6a^2)\,,
\end{equation}
\begin{equation}
	q-\beta_-\Delta = \nu\rho^4 + 6r^2(\alpha^2-a^2) + 6Mr(r^2-\alpha^2)\,,
\end{equation}
where $\nu \equiv \lambda_{2\,lm} + 4$.

More details on how we numerically solve the Schr\"odinger-like wave equation (\ref{eq:schrodinger}) with the potentials Eqs.~(\ref{eq:V0}) to (\ref{eq:V2}) are presented in the \texttt{BlackHawk} manual \cite{BlackHawk}.

\bibliographystyle{apsrev4-1}
\bibliography{biblio}

\end{document}